\documentclass[%
 preprint,
 superscriptaddress,
 amsmath,amssymb,
 aps,
 showkeys,
 titlepage,
 endfloats*.   %
]{revtex4-2}

\usepackage[utf8]{inputenc}
\usepackage[english]{babel}

\usepackage{graphicx}%
\usepackage{dcolumn}%
\usepackage{bm}%

\usepackage[colorlinks = true,
            linkcolor = blue,
            urlcolor  = blue,
            citecolor = red,
            anchorcolor = blue]{hyperref}

\begin{document}

\title{Electric-Field-Dependent Thermal Conductivity in Fresh and Aged Bulk Single Crystalline $\mathrm{BaTiO_3}$} 

\author{Fanghao Zhang}
\affiliation{%
Department of Mechanical Engineering, University of California, Santa Barbara, CA 93106-5070, USA}%

\author{Guanchun Rui}
\affiliation{%
School of Electrical Engineering and Computer Science, Materials Research Institute, The Pennsylvania State University, University Park, PA, 16802 USA}%
\affiliation{%
Arkema Inc., 900 First Avenue, King of Prussia, PA, 19406 USA}%

\author{Yujie Quan}
\affiliation{%
Department of Mechanical Engineering, University of California, Santa Barbara, CA 93106-5070, USA}%

\author{Shantal Adajian}
\affiliation{%
Department of Mechanical Engineering, University of California, Santa Barbara, CA 93106-5070, USA}%

\author{Matthew Delmont}
\affiliation{%
Department of Mechanical Engineering, University of California, Santa Barbara, CA 93106-5070, USA}%

\author{Q. M. Zhang}
\affiliation{%
School of Electrical Engineering and Computer Science, Materials Research Institute, The Pennsylvania State University, University Park, PA, 16802 USA}%

\author{Bolin Liao}
\email{bliao@ucsb.edu} 
\affiliation{%
Department of Mechanical Engineering, University of California, Santa Barbara, CA 93106-5070, USA}%

\date{\today}

\begin{abstract}
Active thermal management requires advances in thermal switching materials, whose thermal conductivity responds to external stimuli.
The electric field, as one of the most convenient and effective stimuli, has shown great potential in tuning the thermal conductivity of ferroelectric materials. 
While previous studies on electric-field-induced ferroelectric thermal switching have primarily focused on thin films and bulk solid solutions with strong extrinsic interface and defect scatterings, bulk single crystals, which can offer clear insights into intrinsic thermal switching mechanisms, have received comparatively less attention.
Here, we demonstrate electric-field-induced thermal switching in bulk single-crystalline BaTiO$_3$ (BTO) at room temperature and elucidate the critical role of domain evolution and aging in governing heat transport.
Using a customized steady-state platform with in-situ electric fields up to ±10 kV/cm, we observe a modulation of thermal conductivity up to 35\% in fresh BTO driven by polarization reorientation and domain restructuring. First-principles finite-temperature lattice-dynamics calculations confirm that the switching behavior primarily originates from anisotropic phonon transport associated with domain configuration rather than strain-induced changes in phonon velocities.
We further reveal that both ambient aging and controlled thermal aging can enhance the switching contrast through the formation and alignment of defect dipoles that modulate phonon-defect scattering. These results establish defect-domain interactions as a powerful design parameter for ferroelectric thermal switches and demonstrate a versatile experimental platform for exploring field-tunable heat transport and phase behavior in bulk functional materials. 
\end{abstract}

                            
\maketitle

\section{Introduction}

With the rapid advancement of artificial intelligence in recent years, the demand for high computational power has increased substantially.
However, the excessive heat generated within chips significantly degrades their performance.
Effective thermal dissipation is crucial for enhancing energy efficiency and preventing operational failures in energy-dense devices and systems~\cite{cheng2025liquid,gebrael2022high}.
Solid-state thermal switching materials, which enable active heat flux control in response to external stimuli, present a promising strategy for advanced thermal management~\cite{wehmeyer2017thermal,li2021transforming}.
An ideal thermal switch should exhibit a high switching ratio ($\kappa_{on}/\kappa_{off}$, defined as the ratio of the maximum to minimum thermal conductivity during switching) along with a rapid response time.
Various external stimuli have been explored to achieve thermal switching through distinct physical mechanisms.
For instance, magnetic fields can induce giant magnetoresistance~\cite{kimling2015spin}, break time-reversal symmetry~\cite{bosisio2015magnetic,yokoyama2011transverse}, and modulate magnon-phonon interactions~\cite{zhao2016heat,zhang2024room}, enabling effective thermal conductivity control, particularly at cryogenic temperatures.
Electrochemical approaches can trigger phase transitions~\cite{lu2020bi} or ion intercalation~\cite{sood2018electrochemical,zhu2016tuning}, yielding high switching ratios but often exhibiting slow response times.
Photoirradiation can induce structural modifications in polymeric materials~\cite{shin2019light}, though efficient photothermally responsive materials remain scarce~\cite{xiang2024high}.
Additionally, pressure~\cite{zhou2022thermal,li2022anomalous} and strain engineering~\cite{chen2023plane} offer pathways to modulate phonon dispersion for thermal switching, yet their integration into practical energy-dense systems remains a challenge.

Among various external stimuli, the electric field is one of the most promising triggers for thermal switching near room temperature due to its ease of application and high-resolution controllability.
Previous studies have demonstrated that ferroelectric materials are promising candidates for electric-field-driven thermal switching~\cite{ihlefeld2015room,liu2019electric,aryana2022observation,liu2023low,wooten2023electric}.
Such thermal switching mechanisms have been achieved through different approaches, including tuning domain wall density in thin-film $\mathrm{Pb(Zr_{0.3}Ti_{0.7})O_3}$ (PZT) with a switching ratio of 1.11~\cite{ihlefeld2015room} and $\mathrm{PbZrO_3}$ (PZO) with a switching ratio of 1.25~\cite{aryana2022observation}, as well as inducing a transition from the antiferroelectric to the ferroelectric phase in thin-film PZO with a switching ratio of 2.2~\cite{liu2023low}.
In contrast, bulk ferroelectric materials exhibit fundamentally different heat transport mechanisms compared to thin films.
In thin-film ferroelectrics, thermal transport is often dominated by phonon scattering at interfaces, domain walls, and grain boundaries, leading to reduced thermal conductivity compared to their bulk counterparts.
Domain wall engineering has been a key mechanism for modulating thermal conductivity in thin-film ferroelectrics, as domain walls act as scattering centers that disrupt phonon propagation, enabling tunable heat transport via electric-field-induced domain reconfiguration~\cite{ihlefeld2015room,aryana2022observation}.
Unlike thin films, where extrinsic boundary scattering plays a dominant role, bulk materials typically experience a stronger influence from intrinsic phonon-phonon scattering, impurity scattering, and polarization-related phonon interactions~\cite{wooten2023electric}.
These differences suggest that bulk single crystalline ferroelectrics can provide a clean platform to explore the intrinsic relationship between thermal conductivity and ferroelectric polarization and the associated strain, whereas in thin films extrinsic effects such as grain boundary scattering and compositional inhomogeneity may obscure the fundamental  relationship between polarization switching and heat transport.
Seminal works by Heremans et al. have demonstrated electric-field-driven thermal switching in bulk ferroelectrics, including commercial $\mathrm{Pb(Zr,Ti)O_3}$ (PZT)~\cite{wooten2023electric} and relaxor $\mathrm{Pb(Mg_{1/3}Nb_{2/3})O_3-33\%PbTiO_3}$(PMN-33PT)~\cite{rashadfar2025electric}. However, both material systems are solid solutions where extrinsic phonon-defect scatterings dominate and have masked the intrinsic thermal switching mechanisms. To this date, a comprehensive understanding of the polarization-thermal conductivity relationship in bulk ferroelectrics remains elusive due to these complex interplay of different phonon scattering mechanisms.

In this study, we systematically investigated the thermal conductivity of bulk single-crystalline ferroelectric barium titanate ($\mathrm{BaTiO_3}$, BTO) near room temperature, where an externally applied electric field enables active thermal switching.
For this purpose, we developed a customized steady-state thermal conductivity measurement system with in-situ electric fields.
Our experimental results show a significant modulation of thermal conductivity with applied electric fields, highlighting the strong coupling between ferroelectric polarization states and phonon transport.
Our first-principles finite-temperature lattice dynamics calculations reveal that the observed thermal switching originates primarily from domain structure evolution and polarization reorientation rather than strain effects.
This finding contrasts with previous assumptions that strain-mediated phonon interactions play a dominant role in the thermal conductivity modulation in ferroelectric materials, emphasizing the importance of polarization engineering in ferroelectric-based thermal switches.
Furthermore, we explored the impact of ferroelectric aging on thermal switching behavior.
Aging, which arises from the gradual alignment of defect dipoles and domain stabilization over time, was found to enhance the variation in thermal conductivity.
Our results demonstrate that aged samples exhibit nearly a 60\% increase in thermal conductivity modulation compared to fresh samples, suggesting that the formation of oriented defect dipoles amplifies the effectiveness of electric-field-induced thermal switching.
This observation underscores the potential of defect engineering as a useful tool for engineering thermal transport properties in ferroelectric materials.

\section{Methods}
\subsection{Steady-state Thermal Conductivity}
To investigate the thermal conductivity of bulk BTO under an in-situ large electric field, we developed a customized steady-state measurement setup.
A schematic representation and a corresponding digital image of the experimental apparatus are shown in Fig.~\ref{fig:fig1}a and Fig.~\ref{fig:fig1}b, respectively.
The single-crystalline BTO samples were purchased from MTI Corporation. 
These samples were cut into plates of dimensions 5 $\times$ 2 $\times$ 0.5 mm$^3$. 
Thermal conductivity measurements were carried out using a steady-state method in a high-vacuum ($10^{-6}$ Torr) environment within the Quantum Design Physical Property Measurement System (PPMS) sample chamber.
A resistive heater (Omega Engineering, 120 $\Omega$ strain gauge) attached to a copper plate functioned as the heat source. 
A copper plate positioned on an Al\textsubscript{2}O\textsubscript{3} plate was affixed to a PPMS puck, acting as both a heat sink and the sample stage.
Type T thermocouples, composed of 25-$\mu$m-diameter copper-constantan wires, were custom-made and used as thermometers.
Stycast or GE varnish, which is electrically insulated but thermally conductive, was employed to establish all contacts. 
Measurements were performed at discrete electric fields between -10 kV/cm and +10 kV/cm. 
We waited for 5 minutes after changing the electric field to achieve thermal equilibrium and eliminate the pyroelectric effect before conducting the measurements. 
LabVIEW was utilized to program the control software.

The thermal conductivity was calculated by:
\begin{equation}
    \kappa = \frac{(IV-Q_\mathrm{rad}-Q_\mathrm{tc})\times L}{A \times \Delta T},
\end{equation}
Where $I$ is the current supplied to the heater, $V$ is the measured voltage across the heater, and $Q_\mathrm{rad}$ and $Q_\mathrm{tc}$ represent the heat loss due to radiation and conduction through the thermocouples, respectively. $A$ is the cross-sectional area of the sample. $L$ and $\Delta T$ are the distance and the temperature difference between the two thermocouples, respectively.
Given that direct measurement of the heat flux is unfeasible, we estimate the net heat conducted through the sample by calculating the power ($IV$) dissipated in the heater resistor and subtracting the losses attributed to radiation and thermal conduction down the thermocouples. 
The radiation heat loss was estimated by:
\begin{equation}
    Q_\mathrm{rad} = \sigma_\mathrm{T} \times (S/2) \times \varepsilon \times (T_\mathrm{hot}^4-T_\mathrm{cold}^4),
\end{equation}
where $\sigma_\mathrm{T}  = 5.67 \times 10^{-8} \ \mathrm{W}\ \mathrm{m}^{-2}\ \mathrm{K}^{-4}$ is the Stefan-Boltzmann constant,
$S$ is the surface area of the sample,
$\varepsilon$ is the infrared emissivity of the radiating surface, and $T_\mathrm{hot/cold}$ are the temperatures measured at the hot and cold ends where the thermocouples are attached. $T_\mathrm{cold}$ is a close approximation for the temperature of the enclosing environment. The factor $\mathrm{\frac{1}{2}}$ in the equation arises from the approximation that, along the length of the sample, the surface temperature varies nearly linearly between $T_\mathrm{hot}$ and $T_\mathrm{cold}$. The total heat losses were calculated to be less than 3\%.
The error bars include the standard deviation from $\geq$ 21 measurements at one electric field, as well as uncertainties from the heater’s current source and the geometry.
It is worth noting that even though the absolute uncertainty is nearly 10\%, mostly due to uncertainty in the sample geometry, the relative error in measuring the change in the thermal conductivity as a function of electric fields is much smaller for the same sample. 
\subsection{Time-domain Thermoreflectance(TDTR) Measurements}
The out-of-plane thermal conductivity of BTO was characterized using time-domain thermoreflectance (TDTR)~\cite{cahill2004analysis, schmidt2008pulse,plunkett2023blending,niyikiza2025thermal}.
Two 100-nanometer-thick aluminum layers were deposited onto the two large surfaces of the sample via electron beam evaporation, and the thickness of the aluminum film was independently validated through picosecond photoacoustic measurements.
The pump beam is modulated at 10 MHz via an electro-optic modulator, inducing periodic heating of the sample surface to facilitate thermoreflectance measurements.
Using a $10\times$ objective lens, the pump and probe beams are focused to $1/e^2$ radii of 35 $\mu$m and 9 $\mu$m, respectively.
During the measurements, the pump and probe beam powers are maintained at approximately 25 mW and 30 mW, respectively, before entering the objective lens.
\subsection{Ferroelectric Measurements}
polarization-electric field (P-E) and strain–electric field (S$_3$-E) hysteresis loops were simultaneously recorded using a modified Sawyer-Tower circuit and a high-voltage amplifier (Trek model 2210). The out-of-plane  strain ((S$_3$-E)) was measured by a linear variable differential transformer (LVDT) fixture linked to a lock-in amplifier (Model SR830 DSP, Stanford Research System). To prevent corona discharge, the samples were immersed in a silicone oil bath at room temperature. A bipolar sinusoidal voltage at 1 Hz was applied to the sample during measurement. 

\subsection{Computational Methods}
To capture the thermal conductivity at room temperature, the temperature-dependent interatomic force constants (IFCs) were extracted by combining ab initio molecular dynamics (AIMD) simulations and the temperature-dependent effective potential (TDEP) technique~\cite{hellman2013temperature,hellman2011lattice}. 
The AIMD simulations were performed within the density functional theory (DFT) framework implemented in the Vienna Ab-initio Simulation Package (VASP) ~\cite{kresse1993ab,kresse1996efficiency,kresse1996efficient}.
The simulations used the projector-augmented wave formalism~\cite{blochl1994projector} with exchange-correlation energy functional parameterized by Perdew, Burke, and Ernzerhof within the generalized gradient approximation ~\cite{perdew1996generalized}. 
Before AIMD simulations, the crystal structure was fully relaxed with energy and Hellmann–Feynman force convergence thresholds of $\rm 10^{-6}~eV$ and $\rm 10^{-4}~eV/\AA$, respectively, and the difference between the optimized lattice constants and the experimental values is within 2\%. The AIMD simulations were performed on a 3×3×3 supercell, 135 atoms in total. 
The electronic self-consistent loop convergence was set to $\rm 10^{-5}~eV$. A single $\rm \Gamma $ point k-mesh with a plane-wave cut-off energy of 450~eV was used to fit the effective energy surface using the TDEP method. 
The simulations were performed at 300~K, with the NVT ensemble using a Nos\'{e}–Hoover thermostat. 
All the simulations were run for 10 ps with a timestep of 2 fs, and the initial 2 ps’s information was discarded due to the non-equilibrium. The q-mesh density, used for thermal conductivity calculations, was set to 10 $\times$ 10 $\times$ 10, with which the values of thermal conductivity were all converged.

\section{Results and Discussions}

\subsection{Thermal Conductivity of Fresh BaTiO$_3$ Single Crystals}
\begin{figure}[!htb]
\includegraphics[width=\textwidth]{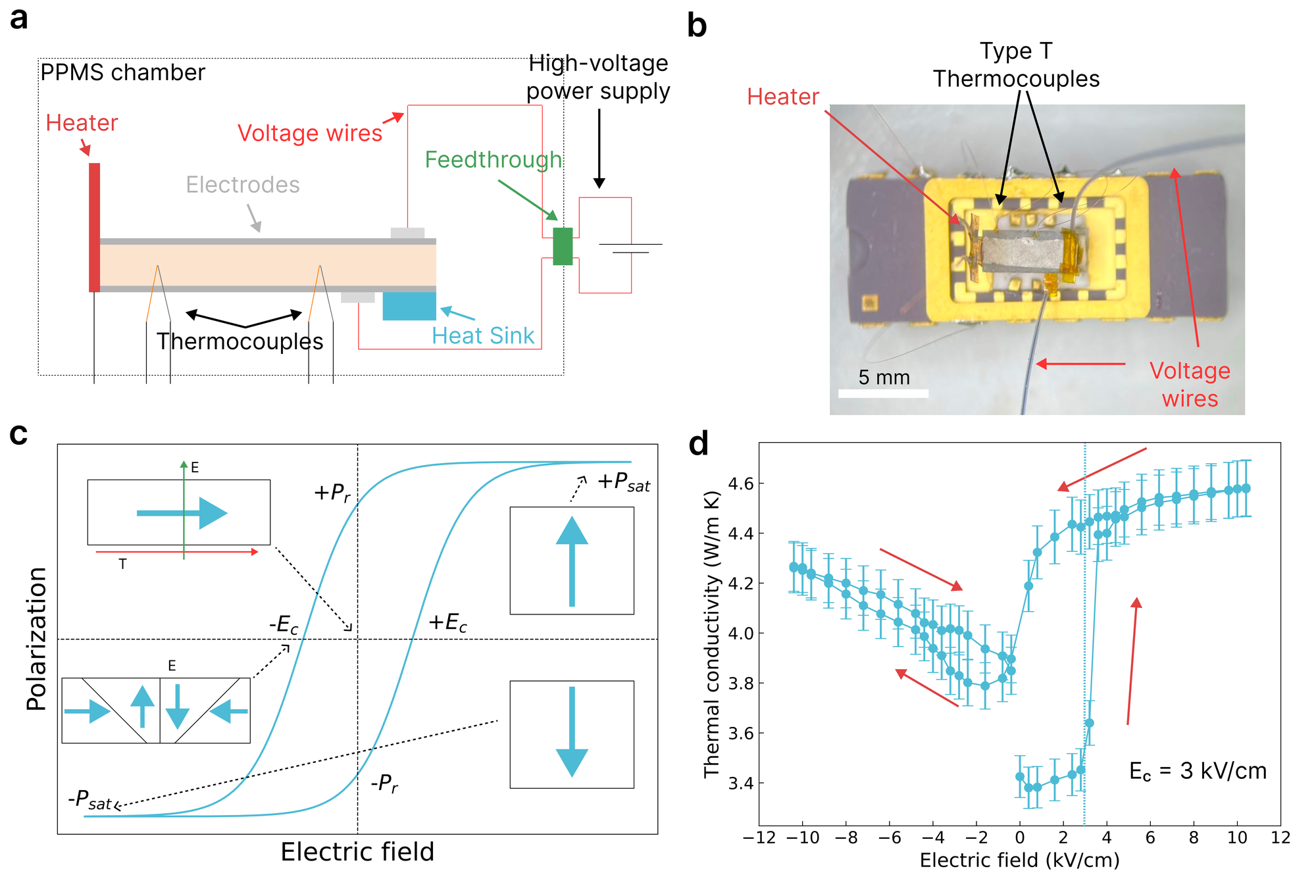}
\caption{\textbf{Steady-state thermal conductivity measurements of fresh BaTiO$_3$ single crystals with in-situ electric fields.} (a) The schematic and (b) a picture of the steady-state experimental setup.  (c) The schematic showing the ferroelectric polarization as a function of electric field in BTO. Here, the polarization represents the out-of-plane component. The temperature gradient is applied along the in-plane direction, while the electric field is applied along the out-of-plane direction.  (d) Thermal conductivity in a fresh BTO single crystal as a function of the external electric field during the first scan. } 
\label{fig:fig1}
\end{figure}

The thermal conductivity measurements were performed on bulk BTO single crystals with an initial net polarization along the (100) crystallographic direction, which is parallel to the in-plane temperature gradient. Because no further poling along the (100) direction was done after we received the sample, the initial state of the sample might contain domains polarized along other directions while the net polarization was along (100).
The electric field was applied in the out-of-plane (001) direction, perpendicular to the temperature gradient.
The polarization of BTO reorients to the out-of-plane direction when the sample is subjected to a sufficiently strong out-of-plane electric field greater than the coercive field $E_c$ = 3 kV/cm, as illustrated in Fig.~\ref{fig:fig1}c. During the polarization switching process, the ferroelectric domain structure evolves, leading to the creation, movement, and annihilation of 90$^{\circ}$ domain walls. The domain size is usually around 0.5 to 1 $\mu$m in single crystal BTO at room temperature~\cite{arlt1990twinning}. As shown in Fig.~\ref{fig:fig1}d, our measured in-plane thermal conductivity of a fresh BTO sample exhibits a minimum around 3.4 W/mK at room temperature before the application of an electric field.
During the first switching cycle, a sharp increase of the thermal conductivity to around 4.4 W/mK was observed when an electric field of 4 kV/cm was applied to BTO, which further increases to 4.6 W/mK with an electric field of 10 kV/cm, resulting in a maximum relative change in the thermal conductivity of approximately 35\%. A hysteretic behavior in the thermal conductivity is observed when the electric field is reduced and later reversed, with a minimum value around 3.8 W/mK with a reversed electric field at -2 kV/cm. This hysteresis is expected since the ferroelectric polarization will not restore to the initial net in-plane configuration during the switching process, when the minimum thermal conductivity corresponds to a domain configuration with an equal number of in-plane and out-of-plane domains and a net zero polarization (Fig.~\ref{fig:fig1}c). Further cycling of the electric field switches the thermal conductivity between 3.8 W/mK and 4.3 W/mK, reducing the switching ratio to around 13\%.

The polarization reorientation in our BTO sample under an applied electric field can influence the measured thermal conductivity ($\kappa$) through multiple mechanisms:
(i) The transition of polarization from the in-plane to the out-of-plane direction alters the thermal transport anisotropy, shifting the measured thermal conductivity from the lower value along the polarization direction ($\kappa_z$) to the higher value perpendicular to the polarization direction ($\kappa_{xy}$)~\cite{feger2024lead}.
(ii) During the polarization switching process, the ferroelectric domain structure also evolves. The movement of 90$^\circ$ ferroelectric domains introduces local strain, which can modify phonon scattering and impact thermal conductivity.
(iii) The reduction in domain-wall density under an applied electric field can potentially suppress phonon scattering, leading to an overall increase in $\kappa$, although this effect is expected to be weak due to the relatively large domain sizes in bulk single-crystalline BTO (0.5 to 1 $\mu$m) compared to its dominant phonon mean free paths (below 100 nm)~\cite{mante1971phonon,wooten2023electric}. 

\begin{figure}[!htb]
\includegraphics[width=\textwidth]{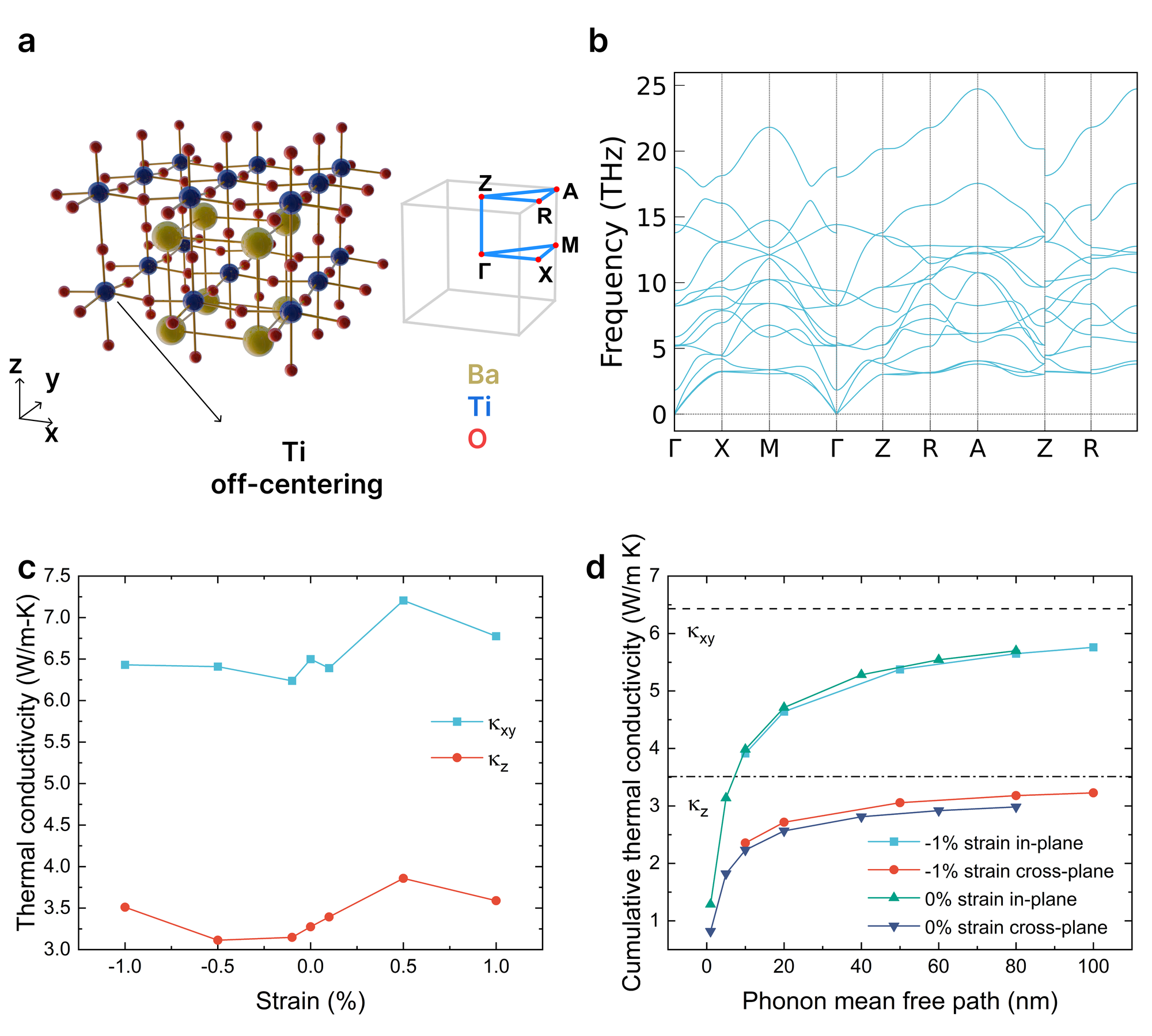}
\caption{\textbf{Calculated thermal conductivity of BTO at 300 K using first-principles finite-temperature lattice dynamics}. (a) The crystal structure of BTO indicating the out-of-plane (z) and in-plane (xy) directions, where Ti off-centering is highlighted. (b) Calculated phonon dispersion of the tetragonal phase of BTO at 300 K, showing a stabilized lattice by phonon anharmonicity. (c) The calculated thermal conductivity along the polarization direction ($\kappa_z$) and perpendicular to the polarization direction ($\kappa_{xy}$) in BaTiO$_3$ at 300 K as a function of strain (-1 to 1\%). A positive (negative) sign indicates a tensile (compressive) strain along the polarization direction. (d) The cumulative thermal conductivity of BTO at 300 K as a function of phonon mean free path.} 
\label{fig:fig2}
\end{figure}

To elucidate the dominant mechanism, we conducted first-principles simulation of the finite-temperature lattice dynamics in BTO at 300 K, taking into account the influences of the anisotropy, the electric-field-induced strain, and the domain size on its thermal conductivity. The results are summarized in Fig.~\ref{fig:fig2}. Fig.~\ref{fig:fig2}a shows the crystal structure of the simulated BTO in its ferroelectric tetragonal phase at 300 K, where the Ti off-centering leading to the polarization is highlighted. Fig.~\ref{fig:fig2}b shows the phonon dispersion of the tetragonal BTO phase at 300 K, where anharmonicity removes imaginary phonons and stabilizes the lattice. Fig.~\ref{fig:fig2}c presents the calculated thermal conductivity of BTO along the polarization direction ($\kappa_z$) and perpendicular to the polarization direction ($\kappa_{xy}$) at 300 K, showing a significant anisotropy at zero strain ($\kappa_{xy}$ = 6.2 W/mK and $\kappa_z$ = 3.3 W/mK). Two previous calculations also showed significant anisotropy in the thermal conductivity of BTO, while the absolute values differ from our results ($\kappa_{xy}$ = 6.2 W/mK and $\kappa_z$ = 4.2 W/mK by Negi et al.~\cite{negi2023thickness} and $\kappa_{xy}$ = 4.5 W/mK and $\kappa_z$ = 3.7 W/mK by F\'{e}ger et al.~\cite{feger2024lead}). The discrepancy may be due to different exchange-correlation functionals and other parameters used in the calculations. Experimentally, previous steady-state measurements of single-crystalline BTO with a mixture of polar domains produced thermal conductivity values around 6 W/mK at 300 K~\cite{suemune1965thermal,mante1967thermal}. Tachibana et al. used the steady-state method to measure single-domain BTO samples and found $\kappa_{xy}$ = 5.5 W/mK and $\kappa_z$ = 4 W/mK at 300 K~\cite{tachibana2008thermal}. Using transient frequency-domain thermoreflectance (FDTR), F\'{e}ger et al. measured $\kappa_{xy}$ = 3.9 W/mK and $\kappa_{z}$ = 2.6 W/mK. Despite the sizable spread of the reported thermal conductivity values, likely due to variations in sample qualities and aging (discussed later), both the simulation and experimental results confirm the strong anisotropy of the thermal conductivity in BTO.

We further simulated the strain effect on the thermal conductivity in ferroelectric tetragonal BTO, as shown in Fig.~\ref{fig:fig2}c. In our simulation, a positive (negative) strain indicates a tensile (compressive) strain along the polarization direction. The most relevant result for our experiment is $\kappa_{xy}$ as a function of the tensile strain along the polarization direction, which corresponds to our measurement when a strong out-of-plane electric field is applied. With the applied electric field in our measurement, the maximum strain achieved in our measurement is estimated to be under 0.5\%~\cite{pan1988field}. Based on the simulation results shown in Fig.~\ref{fig:fig2}c, we expect that the strain effect on thermal conductivity is relatively small in our case and can be responsible for the slight increase of thermal conductivity beyond 4 kV/cm. This is in contrast to a previous investigation of a PZT stack, where the field-induced change in the thermal conductivity was attributed to the strain-dependent acoustic phonon velocity~\cite{wooten2023electric}. 
As shown in Fig.~\ref{fig:fig2}d, our simulations of the cumulative thermal conductivity as a function of phonon mean free path reveal that the dominant heat-carrying phonons in BTO have mean free paths shorter than 100 nm. This is consistent with a previous simulation~\cite{negi2023thickness} and is significantly smaller than the typical domain size in bulk BTO at 300 K. Therefore, domain-wall scattering is expected to contribute negligibly to thermal transport under these conditions.

Given the importance of the thermal conductivity anisotropy, we performed complementary measurements using time-domain thermoreflectance (TDTR) to probe thermal transport along the out-of-plane (001) direction as a function of the out-of-plane electric field. The result is shown in Fig. S1 in the Supplementary Information. The TDTR result showed an opposite trend to the steady-state result shown above: An initial higher thermal conductivity with zero field (4.2 W/mK), which is suppressed by the applied field and reaches a minimum of 2.3 W/mK at 10 kV/cm. Assuming 10 kV/cm nearly saturates the polarization along the (001) direction ~\cite{shieh2009hysteresis}, $\kappa_z$ of a single-domain BTO should approach a minimum value of 2.3 W/mK, as indicated by our TDTR result. This suggests that a thermal conductivity switching from 2.3 W/mK to 4.6 W/mK is possible if the polarization of BTO can be reversibly switched between a fully in-plane state to a fully out-of-plane state, leading to a large switching ratio of 2 at room temperature. In principle, this mode can be achieved by biasing BTO with strain~\cite{burcsu2004large}.

\subsection{Thermal Conductivity of Aged BaTiO$_3$ Single Crystal}
\begin{figure}[!htb]
\includegraphics[width=\textwidth]{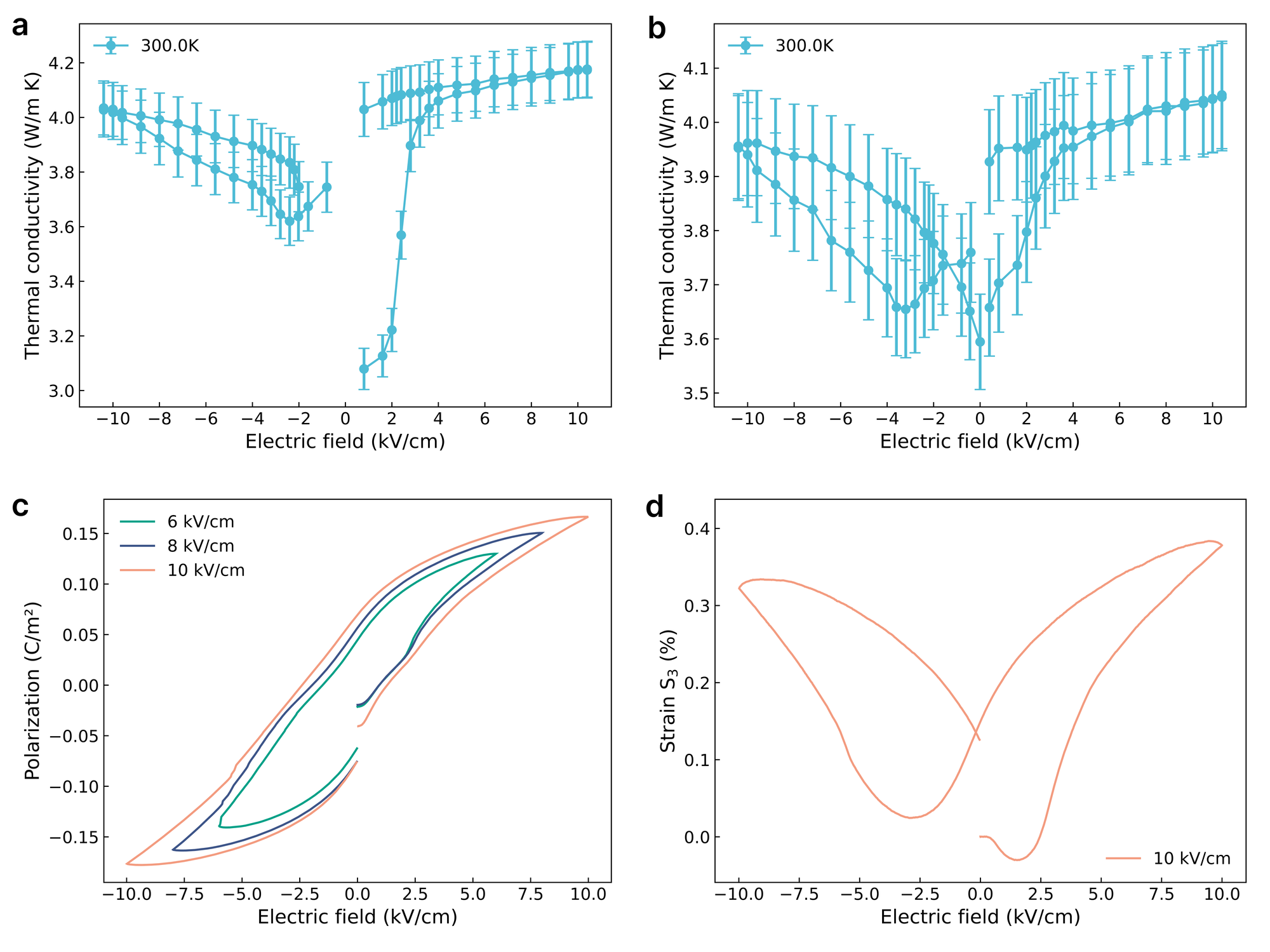}
\caption{\textbf{Electric-field-dependent thermal conductivity in time-aged BaTiO$_3$(100) single crystals}. The thermal conductivity perpendicular to the polarization direction ($\kappa_{xy}$) in BaTiO$_3$ at 300 K as a function of electric fields during (a) the first and (b) the second scan. The initial increase in thermal conductivity during the first scan is attributed to the polarization switch from in-plane to out-of-plane direction. (c)The polarization-electric field (P-E) hysteresis loop and (d) the strain–electric field (S-E) hysteresis loop of the time-aged BaTiO$_3$ single crystal.} 
\label{fig:fig3}
\end{figure}

Aging refers to the gradual change in properties due to the migration of charged defects and is a common phenomenon in ferroelectric materials~\cite{genenko2015mechanisms}. Therefore, in order to develop ferroelectric-based thermal switching applications, the impact of aging on the thermal conductivity of ferroelectric materials is critical.
For this purpose, we investigated the electric-field-dependent thermal conductivity of two BaTiO$_3$ single crystals after going through different aging processes. The first sample was initially poled along the (100) direction and then left under ambient conditions for 4 months. This sample is later referred to as the ``time-aged sample''. After this aging process, the originally transparent sample developed a slight yellow color, possibly due to oxygen loss and the creation of charged oxygen vacancies~\cite{sun2017crystalline}. As shown in Fig.~\ref{fig:fig3}, this sample shows a lower thermal conductivity at zero field than the fresh sample, which could be attributed to the increased defects during aging.
The field-dependence of the thermal conductivity in the time-aged sample displays a trend similar to that of the fresh sample.
A pronounced rise in thermal conductivity occurs at an external field of approximately 2 kV/cm, corresponding to the polarization reorientation from the in-plane to the out-of-plane direction—mirroring the behavior observed for the fresh sample in Fig.~\ref{fig:fig1}d.
Across the entire field range, however, the aged sample exhibits lower thermal conductivity than the fresh one, likely due to defects introduced during aging, which increases phonon–defect scattering and shortens the phonon mean free path.
This sample shows typical ferroelectric polarization-electric-field (P-E) and strain-electric-field (S-E) responses, as shown in Fig.~\ref{fig:fig3}c and Fig.~\ref{fig:fig3}d, with a relatively symmetric strain reaching around 0.4\% at -10 kV/cm.

\begin{figure}[!htb]
\includegraphics[width=\textwidth]{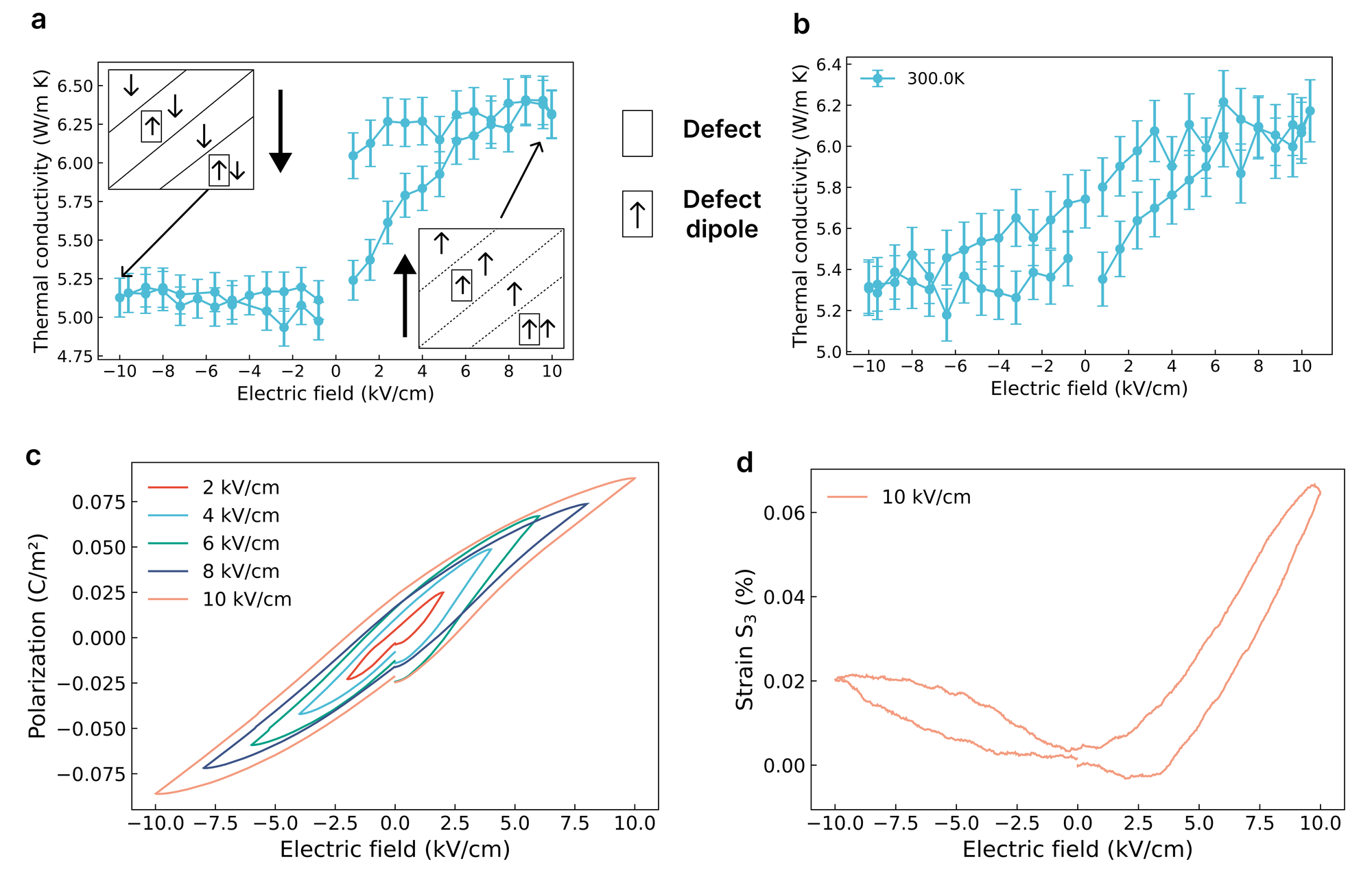}
\caption{\textbf{Electric-field-dependent thermal conductivity in heat-aged BaTiO$_3$ (001) sample.} The thermal conductivity of the heat-aged sample during (a) the first and (b) the second scan of the electric field. The asymmetric thermal switching in the first scan is attributed to the defect dipoles formed during the aging process. (c)The polarization-electric field (P-E) hysteresis loop and (d) the strain–electric field (S-E) hysteresis loop of the heat-aged sample.} 
\label{fig:fig4}
\end{figure}

The second sample was poled along the (001) direction and then underwent a heat treatment at 400 K for 2 hours. This sample is later referred to as the ``heat-aged sample''. At an elevated temperature, charged point defects, such as oxygen vacancies, become mobile and migrate under the influence of the initial polarization. This migration of charged defects forms defect dipoles that are aligned with the initial polarization direction [(001) direction in this case]. Once formed, these defect dipoles are frozen after the temperature is lowered to 300 K and generate an internal electric field that can impact the ferroelectric switching behavior. In particular, the defect dipoles can pin the polarization to the direction parallel to the defect dipoles, making it harder to switch the polarization to the opposite direction ~\cite{sun2005stabilization, cordero2025oxygen}. The formation of defect dipoles has been extensively explored to engineer ferroelectric properties~\cite{ren2004large,lai2023giant} but its consequence on thermal conductivity has not been examined previously. Here, electric-field-dependent thermal conductivity in the heat-aged sample is shown in Fig.~\ref{fig:fig4}. Interestingly, the thermal conductivity of the heat-aged sample at zero field [$\kappa_{xy}$ in this case due to the initial (001) polarization] is around 5 W/m K, which is higher than $\kappa_{xy}$ measured in the fresh BTO sample. This is likely due to the increased order of charged point defects as they form defect dipoles. Notably, 
unlike the symmetric thermal switching behavior under electric fields with opposite polarity in the fresh and time-aged BTO samples, the heat-aged sample showed an asymmetric variation in the thermal conductivity under positive and negative electric fields, as shown in Fig.~\ref{fig:fig4}a during the first scan.
With a positive electric field, the change in thermal conductivity is around 15\% under 5 kV/cm electric field and increases to 25\% under 10 kV/cm (from 5 W/mK to 6.25 W/mK). This switching ratio is largely retained in subsequent switching cycles (as shown in Fig.~\ref{fig:fig4}b), which is almost twice as large as achieved in the fresh BTO sample. With a negative electric field, however, the thermal conductivity shows almost no change with the field strength. This unipolar switching behavior can be explained by the pinning effect of the defect dipoles, which makes it more difficult to switch the ferroelectric domains to the opposite direction as the defect dipoles. 
The thermal conductivity showed hysteresis in the second scan, right after the first scan of the electric field, as shown in Fig.~\ref{fig:fig4}b, while the asymmetric switching behavior is preserved.

The ferroelectric properties of the heat-aged (001) sample are shown in Fig.~\ref{fig:fig4}c and d.
The slight pinching of the P-E loop in the normal mode, as shown in Fig.~\ref{fig:fig4}c, is a typical feature of a ferroelectric aged in a polarized state~\cite{jin2014decoding}.
The asymmetric S-E loop, as shown in Fig.~\ref{fig:fig4}d, highlights the impact of the defect dipoles.
However, the maximum strain we observed in the heat-aged sample is much smaller than the time-aged sample, indicating suppressed domain switching due to defect dipole pinning.
Interestingly, despite the significantly suppressed strain response in heat-aged BTO, the variation in thermal conductivity under an electric field with one polarity is enhanced.
The pronounced hysteresis observed during the second electric field sweep (Fig.~\ref{fig:fig4}d), together with our calculations (Fig.~\ref{fig:fig2}c), indicates that thermal transport in the heat-aged sample is not governed by a macroscopic strain, but rather by electric-field-induced reorientation of defect dipoles, which modulates phonon-defect scattering.
In the zero-field state, the alignment of defect dipoles reduces phonon scattering, leading to a higher baseline thermal conductivity in the heat-aged sample.
When an electric field is applied opposite to the orientation of the defect dipoles (negative polarity), the existing defect dipole alignment remains largely unchanged due to the slow kinetics of ion migration~\cite{lai2023giant}.
This misalignment results in enhanced phonon scattering and consequently minimal change in the thermal conductivity even up to 10 kV/cm.
In contrast, when the electric field is applied along the same direction as the defect dipole alignment (positive polarity), the field further reinforces the ordered dipole configuration, suppressing phonon scattering and resulting in a pronounced increase in thermal conductivity.

\begin{figure*}[!htb]
\includegraphics[width=\textwidth]{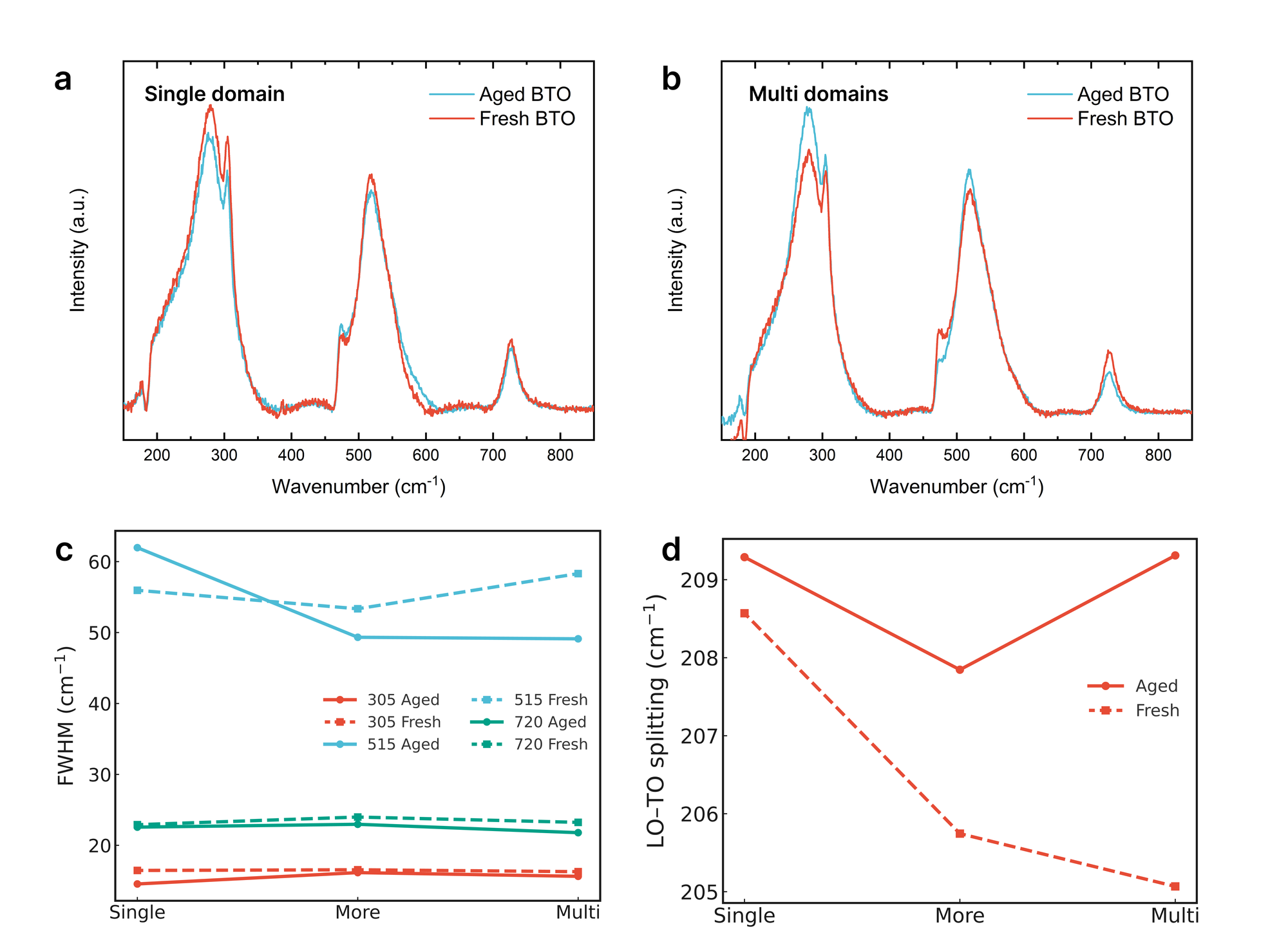}
\caption{\textbf{Raman spectra of fresh and heat-aged BaTiO$_3$.} The Raman spectra after baseline-correction probed from (a) a single ferroelectric domain and (b) multi domains. (c) Full-width-half-maximum (FWHM) of the main peaks and (d) LO-TO splitting probed from a single domain, more domains (two to three domains), and multiple domains.} 
\label{fig:fig5}
\end{figure*}

To provide more evidence to support this hypothesis, the Raman spectra of fresh and heat-aged BTO are collected with different beam sizes to probe a single ferroelectric domain (Fig.~\ref{fig:fig5}a), more domains (two or three domains, Fig.~S2) and multiple domains (Fig.~\ref{fig:fig5}b).
The primary observed peaks are located at 305 $\text{cm}^{-1}$, 515 $\text{cm}^{-1}$ and 720 $\text{cm}^{-1}$.
The 305 $\text{cm}^{-1}$-peak is attributed to the E($\text{TO}_{2}$) phonon mode, the 515 $\text{cm}^{-1}$-peak corresponds to the $\text{A}_{1}$ ($\text{TO}_{3}$) phonon mode and the 720 $\text{cm}^{-1}$-peak can be assigned to the $\text{A}_{1}$ ($\text{LO}_{3}$) phonon mode ~\cite{hermet2009raman}.
The decreased full-width half-maximum (FWHM) in the heat-aged sample (shown in Fig.~\ref{fig:fig5}c), especially when multiple domains are probed, indicates that the alignment of defect dipoles and a reduction of random internal field fluctuations lead to a longer phonon lifetime, which can consequently contribute to a higher thermal conductivity (Fig.~\ref{fig:fig4}a).
In particular, the LO–TO splitting ($\Delta\omega$) in BTO, particularly between the $\text{A}_{1}$($\text{LO}_{3}$) and $\text{A}_{1}$($\text{TO}_{3}$) modes, serves as a direct spectroscopic indicator of the polarization strength and internal electric field~\cite{hermet2009raman}.
The increased $\Delta\omega$ observed in the heat-aged sample reflects the presence and alignment of defect dipoles formed during the aging process.
Moreover, while $\Delta\omega$ remains nearly identical between the fresh and heat-aged samples within single ferroelectric domains, the difference becomes more pronounced when multiple domains are probed.
This behavior suggests that defect dipoles preferentially form or accumulate near domain walls rather than within the bulk of individual domains~\cite{ren2004large,zhang2006aging}.

\section{Conclusion}
In summary, we have demonstrated that electric fields can effectively modulate thermal conductivity in bulk single-crystalline BaTiO$_3$ at room temperature, revealing its potential as a practical ferroelectric thermal switching material. By integrating a custom steady-state thermal measurement platform with in-situ electric fields and complementary TDTR characterization, we identified strong thermal anisotropy and electric-field-driven domain reorientation as key mechanisms behind the observed switching behavior. First-principles finite-temperature lattice-dynamics calculations further showed that the dominant contribution arises from domain-configuration–dependent phonon transport rather than field-induced strain effects. 

Importantly, we uncovered that ferroelectric aging enhances the thermal switching contrast—whether through defect accumulation during long-term ambient aging or through deliberately introduced defect dipoles via thermal aging. These defects stabilize polarization configurations and tune phonon–defect scattering, enabling switching ratios approaching twice those of fresh crystals. Our results highlight aging and defect-dipole engineering as powerful, previously underexplored levers for optimizing ferroelectric thermal switches.

Looking forward, this work opens several new research directions. The strong interplay among defects, domains, and phonon transport suggests that intentional defect-dipole design, domain-wall patterning, and strain-biasing strategies could be combined to realize even larger and more reversible thermal switching at room temperature. Moreover, our experimental platform enables exploration of electric-field-driven phase transitions, nonlinear phonon dynamics, and switching kinetics in a broad range of bulk ferroic materials. Ultimately, these advances may accelerate the development of electrically reconfigurable thermal components—thermal transistors, logic elements, and adaptive heat-flow networks—integrated into next-generation electronic, photonic, and energy-conversion systems. 

\begin{acknowledgments}
The authors acknowledge the help with sample preparation from Dr. Basamat Shaheen. 
The authors used ChatGPT (developed by OpenAI) to edit the manuscript for improved clarity and readability; all scientific content was developed and validated by the authors.
This work is based on research supported by the U.S. Office of Naval Research under the award number N00014-22-1-2262. 
Steady-state transport measurements were done at UCSB Materials Research Laboratory (MRL) Shared Experimental Facilities, which are supported by the NSF MRSEC Program under award number DMR-2308708. 
This work used Stampede2 at Texas Advanced Computing Center (TACC) and Expanse at San Diego Supercomputer Center (SDSC) through allocation MAT200011 from the Advanced Cyberinfrastructure Coordination Ecosystem: Services \& Support (ACCESS) program, which is supported by National Science Foundation grants 2138259, 2138286, 2138307, 2137603, and 2138296. Use was also made of computational facilities purchased with funds from the National Science Foundation (CNS-1725797) and administered by the Center for Scientific Computing (CSC). The CSC is supported by the California NanoSystems Institute and the Materials Research Science and Engineering Center (MRSEC; NSF DMR 2308708) at UCSB. 
\end{acknowledgments}

\bibliography{references.bib}%

\end{document}